\documentclass[a4,notitlepage,12pt]{jedm}

\renewcommand{\arraystretch}{1.5}
\usepackage[utf8]{inputenc}
\usepackage{authblk}
\usepackage{tabularx}
\usepackage{rotating}
\usepackage{enumitem}
\usepackage{url}
\usepackage{amsmath}
\usepackage{amsfonts}
\usepackage{mathtools}
\usepackage{graphicx}
\usepackage{verbatim}
\usepackage{subfig}
\usepackage{array}
\usepackage{tikz}
\usetikzlibrary{matrix,chains,positioning,decorations.pathreplacing,arrows}

\begin{document}

\title{\centering{Machine Learning Based Student Grade Prediction: A Case Study}}
\date{}
\author[*]{Zafar Iqbal}
\author[**]{Junaid Qadir}
\author[*]{Adnan Noor Mian}
\author[*]{Faisal Kamiran}
\affil[*]{Department of Computer Science,}
\affil[**]{Department of Electrical Engineering,}
\affil[ ]{Information Technology University,}
\affil[ ]{Lahore, Pakistan}
\affil[ ]{\textit {\{mscs13039, junaid.qadir, adnan.noor, faisal.kamiran\}@itu.edu.pk}}
\maketitle

\begin{abstract}
In higher educational institutes, many students have to struggle hard to complete different courses since there is no dedicated support offered to students who need special attention in the registered courses. Machine learning techniques can be utilized for students' grades prediction in different courses. Such techniques would help students to improve their performance based on predicted grades and would enable instructors to identify such individuals who might need assistance in the courses. In this paper, we use Collaborative Filtering (CF), Matrix Factorization (MF), and Restricted Boltzmann Machines (RBM) techniques to systematically analyze a real-world data collected from Information Technology University (ITU), Lahore, Pakistan. We evaluate the academic performance of ITU students who got admission in the bachelor's degree program in ITU’s Electrical Engineering department. The RBM technique is found to be better than the other techniques used in predicting the students' performance in the particular course.
\end{abstract}

\section{Introduction}
Since universities are prestigious places of higher education, students' retention in these universities is a matter of high concern \cite{aud2013condition}. It has been found that most of the students' drop-out from the universities during their first year is due to lack of proper support in undergraduate courses  \cite{callender2009part} \cite{macdonald1992meeting}. Due to this reason, the first year of the undergraduate student is referred as a ``make or break'' year. Without getting any support on the course domain and its complexity, it may demotivate a student and can be the cause to withdraw the course. There is a great need to develop an appropriate solution to assist students retention at higher education institutions. Early grade prediction is one of the solutions that have a tendency to monitor students' progress in the degree courses at the University and will lead to improving the students' learning process based on predicted grades.

Using machine learning with Educational Data Mining (EDM) can improve the learning process of students. Different models can be developed to predict students' grades in the enrolled courses, which provide valuable information to facilitate students' retention in those courses. This information can be used to early identify students at-risk based on which a system can suggest the instructors to provide special attention to those students \cite{iraji2012students}. This information can also help in predicting the students' grades in different courses to monitor their performance in a better way that can enhance the students' retention rate of the universities.

Several research studies have been conducted to assess and predict students' performance in the universities. In  \cite{iqbal2016admission}, we analyzed various existing international studies and examined the admission criterion of ITU to found which admission criterion factor can predict the GPA in the first semester at the undergraduate level. From the results, we found that Higher Secondary School Certificate (HSSC) performance and entry test performance are the most significant factors in predicting academic success of the students in the first semester at university. In this study, we are further extending this research and examining the effectiveness of the performance of students of ITU in enrolled courses using machine learning techniques.

In this study, we applied various techniques (CF, SVD, NMF, and RBM) on the real-world data of ITU students. The CF techniques are one of the most popular techniques for predicting students' performance \cite{sarwar1998using}, which works by discovering similar characteristics of users and items in the database; CF, however, does not provide an accurate prediction for a sparse database. The SVD technique makes better predictions as compared to CF algorithms for sparse databases by capturing the hidden latent features in the dataset while avoiding overfitting \cite{berry1995using}. The NMF technique allows meaningful interpretations of the possible hidden features compared to other dimensionality reduction algorithms such as SVD \cite{golub2012matrix}. Finally, RBM can also be used for collaborative filtering and was used for collaborative filtering during the Netflix competition \cite{salakhutdinov2007restricted}. \cite{toscher2010collaborative} tried to use RBM on the KDD Cup dataset and got promising results.

The contributions of this paper are:

\begin{enumerate}
  \item We systematically reviewed the literature about grade/GPA prediction and comprehensively presented them.
  \item We analyzed a real world data collected from 225 undergraduate students of Electrical Engineering Department at ITU.
  \item We evaluated state of the art machine learning techniques (CF, SVD, NMF, and RBM) in predicting the performance of ITU students.
  \item We proposed a feedback model to calculate the student's knowledge for particular course domain and provide feedback if the student needs to put more effort in that course based on the predicted GPA.
  \item We proposed a fitting procedure for hidden Markov model to determine the student performance in a particular course with utilizing the knowledge of course domain.
\end{enumerate}

Rest of the paper is organized as follows. In Section 2, we will describe related work proposed in the literature. Different machine learning techniques that can be utilized to predict students' GPA are briefly outlined in section 3. The methodology of the study for this paper and the performance of the ITU students in different courses are described in Section 4. We present the results and findings of our study in Section 5. We described the insights that hold for our study in Section 6. We highlight some limitations of this study in Section 7. Finally, we conclude the paper in Section 8.

\section{Related Work}
Numerous research studies have been conducted to predict students' academic performance either to facilitate degree planning or to determine students at risk.
\subsection{Matrix Factorization}
\cite{thai2011factorization} proposed matrix factorization models for predicting student performance of Algebra and Bridge to Algebra courses. The factorization techniques are useful in case of sparse data and absence of students' background knowledge and tasks. They split the data into trainset and testset. The data represents the log files of interactions between students and computer aided tutoring systems. \cite{thai2011matrix} extended the research and used tensor-based factorization to predict student success. They formulated the problem of predicting student performance as a recommender system problem and proposed tensor-based factorization techniques to add the temporal effect of student performance. The system saves success/failure logs of students on exercises as they interact with the system.

\subsection{Personalized Multi-Linear Regression models (PLMR)}
Grade prediction accuracy using Matrix Factorization (MF) method degrades when dealing with small sample sizes. \cite{elbadrawy2016predicting} investigated different recommender system techniques to accurately predict the students' next term course grades as well as within the class assessment performance of George Mason University (GMU), University of Minnesota (UMN) and Stanford University (SU). Their study revealed that both Personalized Multi-Linear Regression models (PLMR) and advance Matrix Factorization (MF) techniques could predict next term grades with lower error rate than traditional methods. PLMR was also useful for predicting grades on assessments within a traditional class or online course by incorporating features captured through students' interaction with LMS and MOOC server logs.

\subsection{Regression and Classification Models}
The final grade prediction based on the limited initial data of students and courses is a challenging task because, at the beginning of undergraduate studies, most of the students are motivated and perform well in the first semester but as the time passed there might be a decrease in motivation and performance of the students. \cite{meier2016predicting} proposed an algorithm to predict the final grade of an individual student when the expected accuracy of the prediction is sufficient. The algorithm can be used in both regression and classification settings to predict students' performance in a course and classify them into two groups (the student who perform well and the student who perform poorly). Their study showed that in-class exams were better predictors of the overall performance of a student than the homework assignment. The study also demonstrated that timely prediction of the performance of each student would allow instructors to intervene accordingly. \cite{zimmermann2015model} considered regression models in combination with variable selection and variable aggregation approach to predict the performance of graduate students and their aggregates. They have used a dataset of 171 students from Eidgenössische Technische Hochschule (ETH) Zürich, Switzerland. According to their findings, the undergraduate performance of the students could explain 54\% of the variance in graduate-level performance. By analyzing the structure of the undergraduate program, they assessed a set of students' abilities. Their results can be used as a methodological basis for deriving principle guidelines for admissions committees.

\subsection{Multilayer Perceptron Neural Network}
Educational Data Mining utilizes data mining techniques to discover novel knowledge originating in educational settings \cite{baker2009state}. EDM can be used for decision making in refining repetitive curricula and admission criteria of educational institutions \cite{calders2012introduction}. \cite{saarela2015analysing} applied the EDM approach to analyze the effects of core Computer Science courses and provide novel information for refining repetitive curricula to enhance the success rate of the students. They utilized the historical log file of all the students of the Department of Mathematical Information Technology (DMIT) at the University of Jyväskylä in Finland. They analyzed patterns observed in the historical log file from the student database for enhanced profiling of the core courses and the indication of study skills that support timely and successful graduation. They trained multilayer perceptron neural network model with cross-validation to demonstrate the constructed nonlinear regression model. In their study, they found that the general learning capabilities can better predict the students' success than specific IT skills.

\subsection{Factorization Machines (FM)}
Next term grade prediction methods are developed to predict the grades that a student will obtain in the courses for the next term. \cite{sweeney2015next} developed a system for predicting students' grades using simple baselines and MF-based methods for the dataset of George Mason University (GMU). Their study showed that Factorization Machines (FM) model achieved the lowest prediction error and can be used to predict both cold-start and non-cold-start predictions accurately. In subsequent studies, \cite{sweeney2016next} explored a variety of methods that leverage content features. They used FM, Random Forests (RF), and the Personalized Multi-Linear Regression (PMLR) models to learn patterns from historical transcript data of students along with additional information about the courses and the instructors teaching them. Their study showed that hybrid FM-RF and the PMLR models achieved the lowest prediction error and could be used to predict grades for both new and returning students.

\subsection{Dropout Early Warning System (DEWS)}
Dropout early warning systems help higher education institutions to identify students at risk, and to identify interventions that may help to increase the student retention rate of the institutes. \cite{knowles2015needles} utilized the Wisconsin DEWS approach to predict the student dropout risk. They introduced flexible series of DEWS software modules that can adapt to new data, new algorithms, and new outcome variables to predict the dropout risk as well as impute key predictors.

\subsection{Hidden Markov Model and Bayesian Knowledge Tracing}
Hidden Markov model has been used widely to model student learning. \cite{van2013properties} investigated solutions of hidden Markov model and concluded that the utilization of a maximum likelihood test should be the preferred method for finding parameter values for the hidden Markov Model. \cite{hawkins2014learning} in a separate study developed and analyzed a new fitting procedure for Bayesian Knowledge Tracing and concluded that empirical probabilities had the comparable predictive accuracy to that of expectation maximization.

In Table \ref{table:literature-review}, we have systematically summarized the studies that are related to our study in a comprehensive way to present a big picture of literature. Our work is related to grade prediction systems, recommender systems, and early warning systems within the context of education. In our study, the approach is to use machine learning techniques to predict course grades of students. We used the state of the art techniques that are described and implemented in this section to do a comparative analysis of different techniques that can predict students' GPA in registered courses. We also develop a model that can be used in a tutoring system indicating the weak students in the course to the instructor and providing early warnings to the student if he/she needs to work hard to complete the course.

\begin{table}[!h]
\caption{Systematic Literature Review}
\label{table:literature-review}
\scriptsize\begin{tabularx}{\textwidth}{XXXXX}
\hline
Study & Study Purpose & Dataset & Methods / Techniques & Relevant Findings \\ \hline
\cite{thai2011matrix} & Factorization approaches to predict student performance. & Two real-world datasets from KDD Cup 2010. & Matrix Factorization & MF technique can take slip and guess factors to predict performance.\\
\hline

\cite{thai2011factorization} & 
Matrix factorization models for predicting student performance. & Two real-world datasets from KDD Cup 2010. & Matrix Factorization and Tensor based Factorization & MF techniques are useful for sparse data to predict the performance. \\
\hline

\cite{hawkins2014learning} & Analyze a new fitting procedure for Bayesian Knowledge Tracing. & 1,579 students working on 67 skill-builder problem sets. & Bayesian Knowledge Tracing & Probabilities have accuracy to Expectation Maximization.\\
\hline

\cite{zimmermann2015model} & Predict graduate performance using undergraduate performance. & 171 students data from ETH Zurich. & Regression models. & Third year GPA of undergraduate can predict graduate performance.\\
\hline

\cite{saarela2015analysing} & Analysing students performance using sparse dataset. & Students data of DMIT 2009 - 2013. & Multilayer perceptron neural network & General learning capabilities can predict the students' success.\\
\hline

\cite{sweeney2015next} & Predict students' course grades for the next enrollment term. & 33000 GMU students data of fall 2014. & Factorization Machine & FM model can predict performance with lower prediction error.\\
\hline

\cite{knowles2015needles} & Build a dropout early warning system. & 2006-07 grade 7 cohorts. & Dropout Early Warning Systems (DEWS). & DEWS can predict the dropout risk as well as impute key predictors.\\
\hline

\cite{elbadrawy2016domain} & Investigate the student and course academic features. & 1,700,000 grades from the University of Minnesota. & Collaborative Filtering and Matrix Factorization & Features-based groups make better grade predictions. \\
\hline

\cite{elbadrawy2016predicting} & Predict next term course grades and within-class assessment performance & 30,754 GMU, 14,505 UMN and 13,130 SU students' data. & Personalized Multi-Linear Regression models (PLMR) & PLMR and MF can predict next term grades with lower error. \\
\hline

\cite{sweeney2016next} & Predict students' grades in the courses they will enroll in the next term. & 33000 GMU students data. & Hybrid FM-RF and the PMLR models & Hybrid FM-RF and PMLR methods can predict students' grades. \\
\hline

\cite{meier2016predicting} & Predict grades of individual students in traditional classrooms. & 700 UCLA undergraduate students data. & Regression and classification & In-class evaluations enables timely identification of weak students.\\
\hline

\cite{xumachine} & Machine learning method for predicting student performance. & 1169 UCLA undergraduate students data. & Latent factor method based on course clustering & Latent factor method performs better than benchmark approaches. \\
\hline

\end{tabularx}
\normalsize
\end{table}

\section{Background}
Machine Learning with EDM has gained much more attention in the last few years. Many machine learning techniques, such as collaborative filtering \cite{toscher2010collaborative}, matrix factorization \cite{thai2011multi}, and artificial neural networks \cite{wang2011data} are being used to predict students' GPA or grades. In this section, we will describe these machine learning techniques and how they are being used to predict students' GPA in registered courses within the context of education.

\subsection{Collaborative Filtering}

Collaborative filtering (CF) is one of the most popular recommender system technique to date. In the educational context, the CF algorithms make predictions of GPA by identifying similar students in the dataset. In this method, predictions are made by selecting and aggregating the grades of other students. In particular, there is a list of $m$ students $S = \{s_1, s_2,..., s_m\}$ and a list of $n$ courses $C = \{c_1,c_2,...,c_n\}$. Each student $s_i$ has a list of courses $C_{si}$, which represents student GPA in a course. The task of CF algorithm is to find a student whose GPAs are similar to some other student. User-based Collaborative Filtering (UBCF) is one of the types of collaborative filtering technique. To predict the student GPA in a course, the UBCF algorithm considers similar students that have similar GPA in same courses. The main steps are:

\begin{enumerate}
  \item The algorithm measures how similar each student in the database to the active student by calculating the similarity matrix.
  \item Identify the most similar students by using $k$ nearest neighbors.
  \item Predict the GPA of the course of the active user by aggregating the GPA of that course taken by the most similar students. The aggregation can be a simple mean or weighted average by taking similarity between students into account.
\end{enumerate}

The $k$ nearest neighbour technique is used to select the neighbourhood for the active user $N(a) \subset U$. The average rating of the neighbourhood users is calculated using the equation \ref{eq:aggregation1}, which becomes the predicted rating for the active use. The grade prediction becomes extremely challenging for the student with a few courses attended which is a well-known drawback of CF technique over the sparse dataset.

\begin{equation}
\hat{r}_{aj}=\frac{1}{|N(a)|} \sum_{i \in N(a)} r_{ij}
\label{eq:aggregation1}
\end{equation}

\subsection{Matrix Factorization}
Matrix factorization is a decomposition of a matrix into two or more matrices. Matrix factorization techniques are used to discover hidden latent factors and to predict missing values of the matrix. In our study, we formulated the problem of predicting student performance as a recommender system problem and used matrix factorization methods (SVD and NMF) which are the most effective approaches in recommender systems.

\subsubsection{Singular Value Decomposition}
Singular Value Decomposition (SVD) is a matrix factorization technique that decomposes students-courses matrix $R$ into  

\begin{equation}
    R = U \Sigma V^T,
\end{equation}

where;

\begin{itemize}
  \item $U$ is an $m \times r$ orthogonal matrix, where $m$ represents number of users and $r$ represents the rank of the matrix $R$,
  \item $\Sigma$ is an $r \times r$ diagonal matrix with singular values along the main diagonal entries and zero everywhere else,
  \item $V$ is an $r \times n$ orthogonal matrix where $n$ represents the number of courses.
\end{itemize}

\begin{figure}[h]
    \centering
    \includegraphics[width=0.9\textwidth]{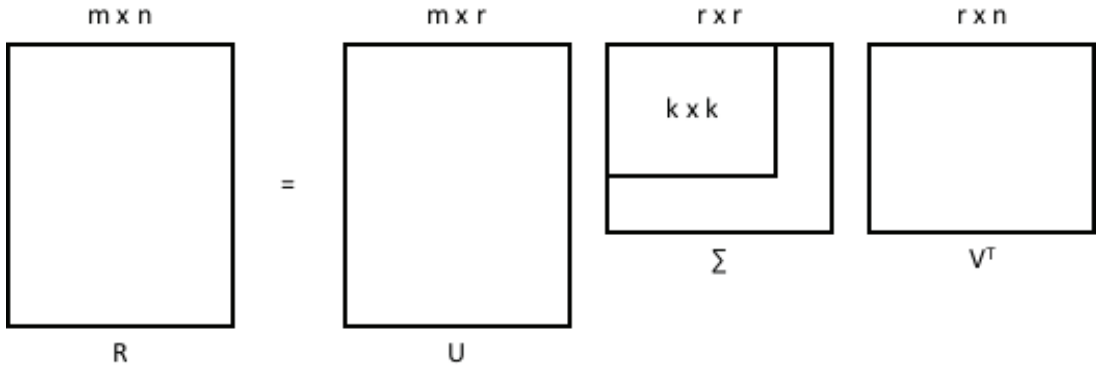}
    \caption{Decomposition of Matrix $R$ by SVD}
    \label{fig:svd}
\end{figure}

The graphical representation of SVD is shown in Figure \ref{fig:svd}. In newly constructed matrices, $r$ represents the rank of the matrix $R$. The values in the matrix $\Sigma$ are known as singular values $\sigma{_i}$, and they are stored in decreasing order of their magnitude. Each singular value $\sigma{_i}$ of the matrix $\Sigma$ represents hidden latent features, and their weights have variance on the values of matrix $R$. The sum of all elements represents the total variance of matrix $R$.

SVD is widely being used to find the best $k$-rank approximation for the matrix $R$. The rank $r$ can be reduced to $k$, where $k < r$, by taking only the largest singular value $k$ which is the first diagonal value of the matrix $\Sigma$ and then reduce both $U$ and $V$ accordingly. The obtained result is a $k$-rank approximation $R_k = U_k \Sigma_k V_k ^T$  of the matrix $R$, in such a way that the Frobenius norm of $R - R_k$ is minimized. The Frobenius norm $(|| R - R_k ||F)$ is defined as simply the sum of squares of elements in $R - R_k$ \cite{deerwester1990indexing}. To make a prediction of the GPA in a course, SVD assumes that each student grade is composed of the sum of preferences of the various latent factors of the courses. To predict the grade of a student $i$ for course $j$ is as simple as taking the dot product of vector $i$ in the student feature matrix and the vector $j$ in the course feature matrix.

The problem with SVD is that it is not effective on big and sparse datasets. Simon Funk proposed to use a Stochastic Gradient Descent (SGD) algorithm to compute the best rank-$k$ matrix approximation using only the known ratings of original matrix \cite{netflix2006}. Stochastic Gradient Descent (SGD) is a convex optimization technique that gets the most accurate values of those two featured matrices that are obtained during the decomposition of the original matrix in the method of SVD. SGD has following steps:

\begin{enumerate}
  \item Re-construct the target students-courses matrix by multiplying the two lower-ranked matrices.
  \item Get the difference between the target matrix and the generated matrix.
  \item Adjust the values of the two lower-ranked matrices by distributing the difference to each matrix according to their contribution to the product target matrix.
\end{enumerate}

Above is a repeated process till the difference is lower than a preset threshold.  By reducing the dimensionality of the students-courses matrix, the execution speed is reduced, and the accuracy of the prediction is increased because of considering only the courses that contribute to the reduced data. Dimensionality reduction leads to the reduction of noise and over-fitting. This method is also used in recommender systems for the Netflix challenge \cite{koren2009matrix}.

\subsubsection{Non-Negative Matrix Factorization}
Non-negative matrix factorization (NMF) is a matrix factorization technique that decomposes a matrix $V$ into two non-negative factor matrices $W$ and $H$ such that

\begin{equation}
    V \approx W H,
\end{equation}

where;

\begin{itemize}
  \item W is a $u \times k$ orthogonal matrix,
  \item H is a $k \times v$ orthogonal matrix.
\end{itemize}

\begin{figure}[!h]
    \centering
    \includegraphics[width=0.6\textwidth]{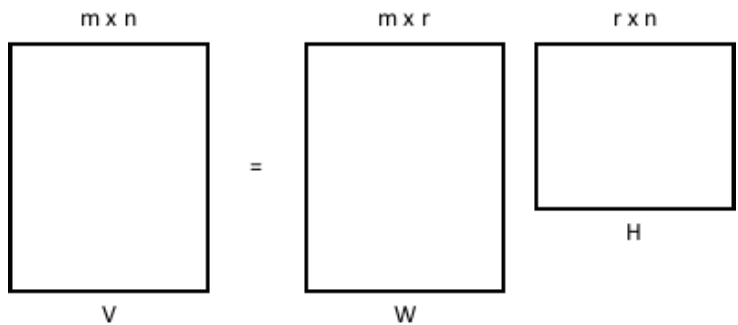}
    \caption{Decomposition of Matrix $V$ by NMF}
    \label{fig:nmf}
\end{figure}

Graphical representation of NMF is shown in Figure \ref{fig:nmf}. NMF is a powerful technique that uncovers the latent hidden features in a dataset and provides a non-negative representation of data \cite{koren2009matrix}. The problem with NMF is to find $W$ and $H$ when the dataset is large and sparse. A sequential coordinate-wise descent (SCD) algorithm can be used with NMF to impute the missing values \cite{franc2005sequential}. NMF imputation using SCD takes all entries into account when imputing a single missing entry.

\subsection{Restricted Boltzmann Machines}
\begin{figure}[h]
\centering
\begin{tikzpicture}[
plain/.style={
  draw=none,
  fill=none,
  },
net/.style={
  matrix of nodes,
  nodes={
    draw,
    circle,
    inner sep=11pt
    },
  nodes in empty cells,
  column sep=0pt,
  row sep=-1cm
  },
>=latex
]

\matrix[net] (mat)
{
|[plain]| \parbox{1.3cm}{\centering Hidden Nodes $h_j$}
&|[plain]| &|[plain]| & & |[plain]| & & |[plain]| & & |[plain]| &  \\
|[plain]| \parbox{1.3cm}{\centering Weights $w_{ij}$}
     &|[plain]| &|[plain]| &|[plain]| &|[plain]| &
    |[plain]| &|[plain]| &|[plain]| &|[plain]| &|[plain]| \\
|[plain]| \parbox{1.3cm}{\centering Visible Nodes $v_i$} & 
    |[plain]| & & |[plain]| & & |[plain]| & & |[plain]| & & |[plain]| &\\
};
\foreach \ai [count=\mi ]in {3,5,7,9,11}
  \draw[<-] (mat-3-\ai) -- node[right] {Course \mi} +(0cm,-1.3);
\foreach \ai in {4,6,8,10}
{\foreach \aii in {3,5,7,9,11}
  \draw[-] (mat-1-\ai) -- (mat-3-\aii);
}
\end{tikzpicture}

\caption{\label{fig:rbm}
A Restricted Boltzmann Machines (RBM) with five courses and four hidden nodes for a specific student.
}
\end{figure}
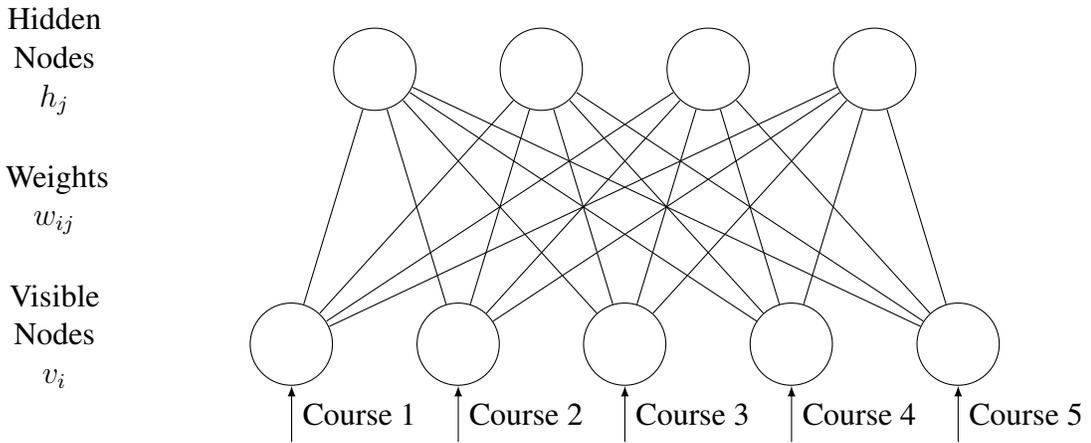

The method of Restricted Boltzmann Machines (RBM) is an unsupervised machine learning method. Unsupervised algorithms are used to find the structural patterns within the dataset. We have used RBM to predict the students' performance in different courses. An RBM is in the form of a bipartite graph that creates two layers of nodes. The first layer is called the visible layer, which contains the input data (Course Grades). These nodes are connected to the second layer which is called the hidden layer that contains symmetrically weighted connections. From the Figure \ref{fig:rbm} we can see that the graph have five visible nodes (Course Grades) denoted by $v_i$ and four hidden nodes indicated by $h_j$. The weights between the two nodes are $w_{ij}$. Here each visible node $v_i$ represents the grade for course $i$, for a particular student.

RBM is a form of Markov Random Field (MRF). MRF is a type of probabilistic model that encodes the structure of the model as an undirected graph for which the energy function is linear in its free parameters. The energy function $E(v,h)$ for RBM can be calculated using the equation \ref{eq:energy}.

\begin{equation}
\label{eq:energy}
    E(v,h) = - a^Tv - b^Th - h^T Wv
\end{equation}   

In the above equation, $W$ represents the weights between the hidden and visible nodes and $a$, $b$ are the offsets of the visible and hidden layers respectively. The probability distributions $P(v,h)$ of visible and/or hidden nodes can be calculated using the equation \ref{eq:distribution}.

\begin{equation}
\label{eq:distribution}
    P(v,h) = 
        \frac{1}
        {Z}e^{-E(v,h)}
\end{equation}

Where $Z$ is a partition function that defines normalization of the distribution. To predict a student grade, one can include an additional visible node $v_p$, for which the value is unknown, but it can be determined by using the energy function given in the equation \ref{eq:prediction}.

\begin{equation}
\label{eq:prediction}
    P(v_p|v,h) \propto
        \frac{1}
        {Z}e^{-E(v_p,v,h)}
\end{equation}

\section{Methods}
We used CF (UBCF), MF (SVD and NMF) and RBM techniques to predict GPA of the student for the courses. A feedback model is developed based on the predicted GPA of the student in a course.

\subsection{Dataset Description}
A real world student data is collected from Electrical Engineering Department at ITU across students of the batch (2013, 2014, 2015). The dataset contains data of 225 undergraduate students enrolled in the Electrical Engineering program. The data of each student contains the students pre-university traits (secondary school percentage, high school percentage, entry test scores and interview), the course credits and the obtained grades of 24 different courses that the students take in different semesters. We consider only letter-grade courses but not fail courses. The information of courses and their domain is shown in Table \ref{table:course-domain}, which was obtained from the curriculum for Electrical Engineering designed for Pakistani Universities \cite{higher2012hec}.

\renewcommand{\arraystretch}{1.2}
\begin{table}[!h]
\caption{Courses Domain Table}
\label{table:course-domain}
\begin{tabular}{l p{3.7in}}
\hline
Course Domain & Courses \\
\hline
Humanities & Communication Skills I, Communication Skills II, Islamic Studies \\
Management Sciences & Industrial Chemistry, Entrepreneurship,  D Lab \\
Natural Sciences & Linear Algebra, Calculus and Analytical Geometry, Complex Variables and Transforms, Probability \& Statistics \\
Computing & Object Oriented Programming, Computing Fundamentals and Programming \\
Electrical Engineering Foundation & Linear Circuit Analysis, Electricity and Magnetism, Electronics Workbench, Electronic Devices and Circuits, Digital Logic Design, Electrical Network Analysis, Electronic Circuit and Design, Signals \& Systems \\
Electrical Engineering Core & Solid State Electronics, Microcontrollers and Interfacing, Electrical Machines, Power Electronics \\
\hline
\end{tabular}
\end{table}

\subsection{Problem Formulation}
For this study, we would like to predict student GPA from the scale 0.0 - 4.0. The given data we have is $\langle Student, Course, GPA \rangle$ triplet and we need to predict GPA for each student for the courses he/she will enroll in the future. In general, we have $n$ students and $m$ courses, comprising an $n \times m$ sparse GPA matrix $G$, where $\{G_{ij} \in R \mid G_{ij} \leq 4\}$  is the grade student $i$ earned in course $j$.

\begin{figure}[!h]
    \centering
    \includegraphics[width=0.7\textwidth]{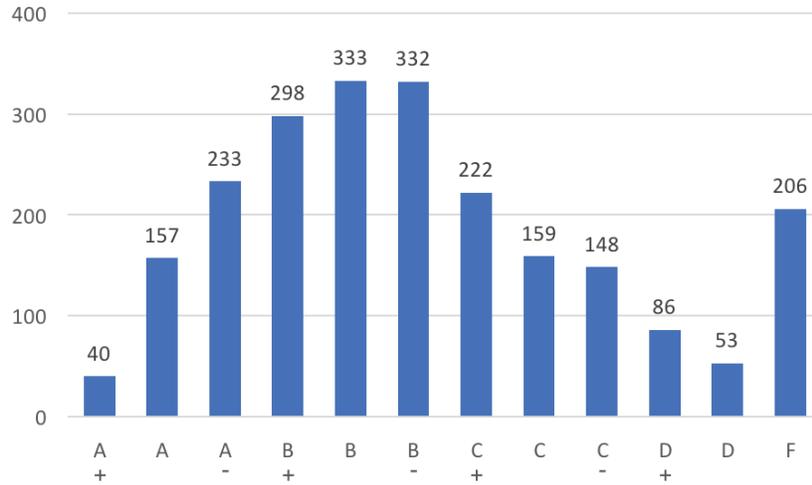}
    \caption{Distribution of students' grades received for the taken courses}
    \label{fig:grades-distribution}
\end{figure}

For training machine learning models, students grades need to be converted to GPA. These grades are converted to numerical GPA values using the ITU grading policy on a 4 point GPA scale with respect to the letter grades A+=4, A=4, A-=3.67, B+=3.33, B=3.0, B-=2.67, C+=2.33, C=2.0, C-=1.67, D+=1.33, D-=1.0 and F=0.0. Figure \ref{fig:grades-distribution} shows the frequency distribution of grades for the students whose grades are available in the dataset. We can see most of the students have B or B- grades in the courses they have taken.

As prediction algorithm works best with centering predictor variables, so all the data were transformed by centering (average GPA of a course is subtracted from all GPAs of that course). The main characteristics of the dataset are shown in the Table \ref{table:exploratoryanalysis}.

\begin{table}[!h]
\centering
\caption{Description of ITU dataset used in this study}
\label{table:exploratoryanalysis}
\begin{tabularx}{\textwidth}{X r}
\hline
Characteristic & Number \\
\hline
Total students & 225 \\
Total courses & 24 \\
Total cells & 5400 \\
Elements (grades) available & 1736 \\
Elements (grades) missing & 3664 \\
Matrix density & 32.14\% \\
\hline
\end{tabularx}
\end{table}

\subsection{Prediction of Student Grades}
As our objective is to predict students GPA in the courses for which he/she needs to enroll in the future, we used CF (UBCF), MF (SVD and NMF) and RBM techniques to predict courses GPA of students. We take the data into a matrix in the form of $\langle Student, Course, GPA \rangle$ triplet. For illustration, here we have taken a few students and courses to display their grades. In the Table \ref{table:students-courses} we can see that a student with Id. SB145 have a GPA 3.67 in the course Electronic Circuit and Design and have a GPA of 4.0 in the D-Lab course. While this student needs to enroll into Linear Circuit Analysis, Islamic Studies, and Signals and System. A student with Id. SB185 have similar GPA in Electronic Circuit and Design course like the student with Id. SB145 and this student need to enroll into Linear Circuit Analysis, Islamic Studies, Signals and Systems, and D-Lab courses.

\begin{table}[h]
\centering
\caption{Students-Courses matrix with students' GPA in particular courses}
\label{table:students-courses}
\begin{tabularx}{\textwidth}{XXXXXX}
\hline
Student Id. & LCA & ECD & IS & SS & DL \\
\hline
SB145 & & 3.67 & & & 4 \\
SB161 & & 4 & & & 3.67 \\
SB185 & & 3.67 & & & \\
SB229 & & & & & \\
SB304 & 2 & & 2.67 & & \\
\hline
\end{tabularx}
\begin{tabular}{lll}
Linear Circuit Analysis (LCA) &  Electronic Circuit and Design (ECD) & Islamic Studies (IS) \\
Signals \& Systems (SS) & D-Lab (DL) & \\
\end{tabular}
\end{table}

\textbf{Collaborative Filtering:} We have used UBCF to predict the students' grades in courses. UBCF do grade prediction of a student $s$ in a course $c$ by identifying student grades in same courses as $s$. For prediction of grades, the neighborhood students $ns$ similar to student $s$ are selected that have taken at least $nc$ courses that were taken by student $s$. To apply UBCF model we first converted the students-courses matrix $R$ into a real-valued rating matrix having student GPA from 0 to 4. To measure the accuracy of this model we have split the data into 70\% trainset and 30\% testset. In UBCF model The similarity between students and courses is calculated using $k$ nearest neighbors.

\textbf{Matrix Factorization:} Matrix factorization is the decomposition of a matrix $V$ into the product of two matrices $W$ and $H$, i.e. $V \approx WH^T$ \cite{koren2009matrix}. In this study, we have used SVD and NMF matrix factorization techniques to predict the student GPA. The main issue of MF techniques is to find out the optimized value of matrix cells for $W$ and $H$.

In \textbf{SVD} approach, the students' dataset is converted into real-valued rating matrix having student grades from 0 to 4. The dataset is split into 70\% for training the model and 30\% for testing the model accuracy. We used Funk SVD to predict GPA in the courses for which the students are shown in Table \ref{table:students-courses} have not yet taken the courses. The largest ten singular values are 191.8012, 18.8545, 14.7946, 13.8048, 12.4328, 11.8258, 11.1058, 10.2583, 9.5020 and 9.1835. It can be observed from the Figure \ref{fig:svd} that the distribution of the singular values of students-courses matrix diminishes quite fast suggesting that the matrix can be approximated by a low-rank matrix with high accuracy. This encourages the adoption of low-rank matrix completion methods for solving our grade/GPA prediction problem.

\begin{figure}[!h]
    \centering
    \includegraphics[width=.75\linewidth]{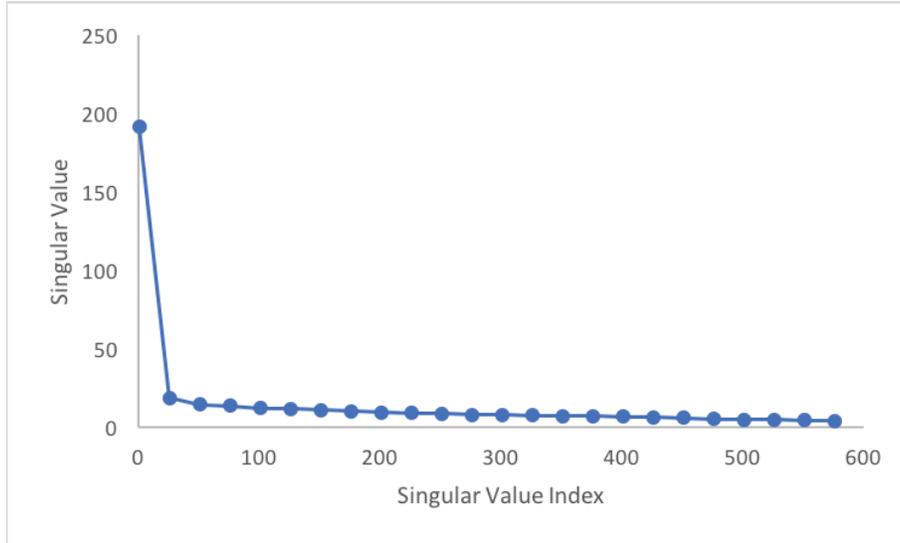}
    \caption{Singular vales distribution of students-courses matrix}
    \label{fig:svd}
\end{figure}

By applying Funk's proposed heuristic search technique called Stochastic Gradient Descent (SGD) gradient to the matrix G we obtained two matrices student and courses dimensional spaces (with the number of hidden features set to two, so as to ease the task of visualizing the data). The stochastic gradient descent technique estimates the best approximation matrix of the problem using greedy improvement approach \cite{pelanek2015student}.

Table \ref{table:student-features} represents the students' features dimensional space, and Table \ref{table:course-features} represents courses' features dimensional space. With the dot product of these features dimensional space we can predict GPA in the courses for which the students are shown in Table \ref{table:students-courses} needs to enroll. Please note that we usually do not know the exact meaning of the values of these two-dimensional space, we are just interested in finding the correlation between the vectors in that dimensional space. For understanding, take an example of a movie recommender system. After matrix factorization, each user and each movie are represented by two-dimensional space. The values of the dimensional space represent the genre, amount of action involved, quality of performers or any other concept. Even if we do not know what these values represent, but we can find the correlation between users and movies using the values of dimensional space.

\renewcommand{\arraystretch}{1.5}
\begin{table}[h!]
\centering
\setlength\tabcolsep{4pt}
\begin{minipage}{0.48\textwidth}
\centering
\caption{Students' features dimensional space} 
\label{table:student-features}
\begin{tabularx}{\textwidth}{Xrr}
    \hline
    Name & V1 & V2 \\ 
    \hline
    SB145 & 0.39 & 0.18 \\ 
    SB161 & 0.45 & 0.20 \\ 
    SB185 & 0.42 & 0.20 \\ 
    SB229 & -0.31 & 0.02 \\ 
    SB304 & 0.09 & 0.12 \\ 
    \hline
    \end{tabularx}
\end{minipage}%
\hfill
\begin{minipage}{0.48\textwidth}
\centering
\caption{Courses' features dimensional space}
\label{table:course-features}
    \begin{tabularx}{\textwidth}{Xrr}
        \hline
        Name & V1 & V2 \\ 
        \hline
        Linear Circuit Analysis & 1.19 & -0.04 \\ 
        Electronic Circuit and Design & 0.94 & 0.10 \\ 
        Islamic Studies & 1.77 & -0.03 \\ 
        Signals and Systems & 0.34 & 0.20 \\  
        D-Lab & 0.46 & 0.18 \\ 
        \hline
    \end{tabularx}
\end{minipage}
\end{table}

\begin{figure}[!h]
    \centering
    \includegraphics[width=.75\linewidth]{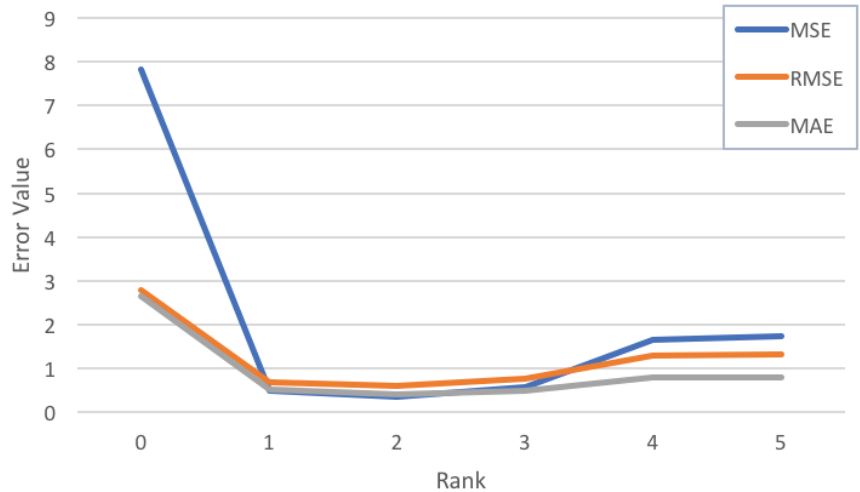}
    \caption{Rank-$k$ using NMF}
    \label{fig:nmf-rank}
\end{figure}

In \textbf{NMF} approach, we have a $u \times v$ matrix $V$ with non-negative entries of student grades from 0 - 4 that decomposes into two non-negative, rank-$k$ matrices $W (u \times k)$ and $H (k \times v)$ such that $V \approx WH$. Before decomposing a matrix into two matrices first, we need to choose a rank-$k$ for NMF that gives the smallest error for grade predictions of the students-courses matrix. In our experiments with NMF, the rank-$k$ 2 gives the minimum Mean Squared Error (MSE) as shown in the Figure \ref{fig:nmf-rank}. So, we have used two as rank-$k$ value and decomposed the matrix into $W$ and $H$.

\textbf{Restricted Boltzmann Machines:}
We have also used RBM an unsupervised learning technique to predict the student grades in different courses. RBM has been used to fill the missing data in a students-courses matrix. We have split the data into 70\% trainset and 30\% testset. We have trained the RBM method with a learning rate of 0.1, momentum constant of 0.9, the batch size of 180, and for 1000 epochs.

\subsection{Feedback Methodology}
\begin{figure}[h]
    \centering
    \includegraphics[width=0.99\textwidth]{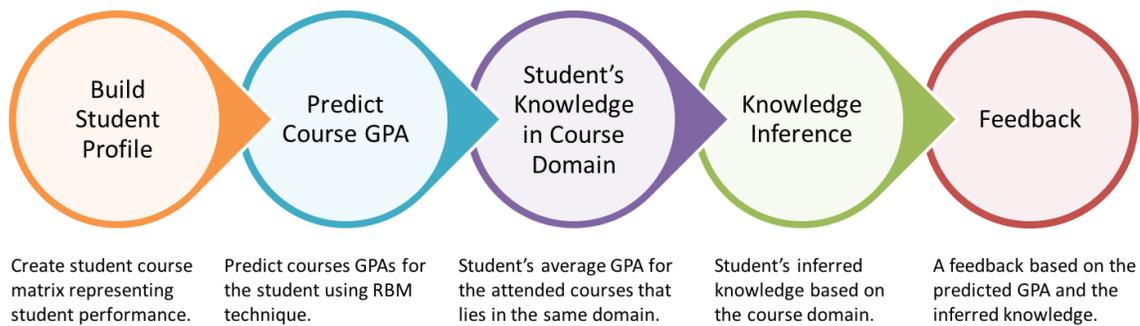}
    \caption{Main steps of feedback model}
    \label{fig:flow-diagram}
\end{figure}

Machine learning techniques can be utilized to identify the weak students who need appropriate counseling/advising in the courses, by early predicting the courses grades. A feedback model
that we have developed will calculate the student's knowledge in the particular course domain based on the results it gives feedback to the instructor about the courses in which a student is weak. The detail of the feedback model is given below and represented in a Figure \ref{fig:flow-diagram}.

\begin{enumerate}
  \item \textbf{Build Student Profile:} In the first phase of feedback model; we have to parse students and courses data into the form of $\langle Student, Course, GPA \rangle$ triplet to built students' profile. A students-courses matrix $R$ is created that contains students' performance in each course taken. In a matrix $R$, students are represented in rows and courses are represented in columns. The value of each cell of matrix $R$ is $R_{ij}$, that can be calculated using the equation \ref{matrix-entry-equation}.

\begin{equation}\
\label{matrix-entry-equation}
R_{ij} =   \left \{
  \begin{tabular}{lr}
  student's $i$ mark on course $j$, & if the student enrolled in course $j$ \\
  empty, & if the student did not enroll in course $j$ \\
  \end{tabular}
\right \}
\end{equation}

For the courses in which a student did not enroll, $R_{ij}$ will be empty. For illustration, a small chunk of the dataset is presented in matrix given below. This matrix holds the dataset of five different students and five different courses.

\begin{equation}\
\label{students-courses-matrix}
R_{ij} = \left \{
  \begin{tabular}{llllll}
   & & 3.67 & & & 4 \\
   & & 4 & & & 3.67 \\
   & & 3.67 & & & \\
   & & & & & \\
   & 2 & & 2.67 & & \\
   \end{tabular}
\right \}
\end{equation}

\item \textbf{Predict Course GPA:}
Now we have a matrix $R$, for which we are interested to find the unknown GPAs for the courses, which the student has not taken yet. To find the predicted GPA we have used CF (UBCF), MF (SVD and NMF), and RBM techniques. Detailed methodology for these techniques is described in section 4.

\item \textbf{Students' Knowledge in Course Domain:}
In our feedback model, student knowledge in different course domains is calculated by taking an average of GPAs for the courses the student has taken which fall into the same domain by using the course domain table (Table \ref{table:course-domain}).

\item \textbf{Knowledge Inference:}
Hidden Markov Model (HMM) is a model used to predict students' performance based on their historical performance. According to the model, the probability of knowledge $P(L_j)$ increases with every step $j$ and can be calculated using the equation \ref{bkt-equation}.

\begin{equation}\
\label{bkt-equation}
P(L_j) = P(L_{j-1}) + P(T) (1 - P(L_{j-1})),
\end{equation}

where;

\begin{itemize}
  \item $P(L_j)$ is the probability of knowledge in the step $j$,
  \item $P(L_{j-1})$ is the probability of knowledge in the previous step,
  \item $P(T)$ is the probability of learning,
  \item $(1 - P(L_{j-1})$ is the knowledge that is unknown.
\end{itemize}

Using the equation \ref{bkt-equation}, student knowledge is measured by inferring his knowledge in the course domain. As we know the probability of the knowledge in the previous step is the predicted GPA for the student in the subject. To calculate the knowledge gain course domain average has been converted into the range (0 to 1) and multiplied by the learning rate 0.005.

\item \textbf{Feedback:}
After computing the student knowledge in particular course domain and knowledge inference, the feedback is made. If the student knowledge inference results are less than 2.67 GPA in a course, then the system generates a warning that the student needs effort in that course. In this way, feedback results can inform the instructors that the student is weak in a particular course.
\end{enumerate}

\section{Results}

\subsection{Correlation Analysis}
\begin{figure*}[!h]
\centering
\begin{minipage}[b]{.45\textwidth}
\includegraphics[width=.99\linewidth]{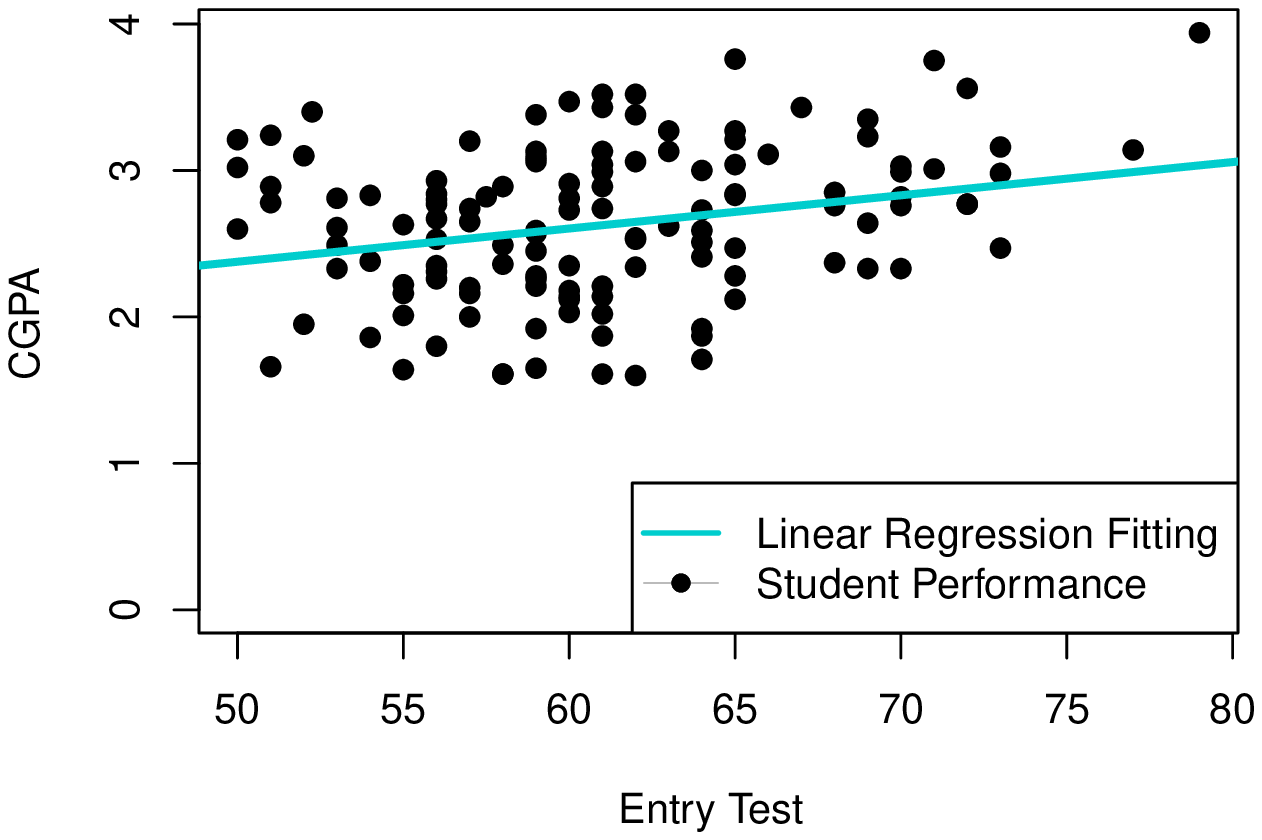}
\caption{Correlation between entry test and CGPA}
\label{fig:linear-regression-fitting-2}
\end{minipage}\qquad
\begin{minipage}[b]{.45\textwidth}
\includegraphics[width=.99\linewidth]{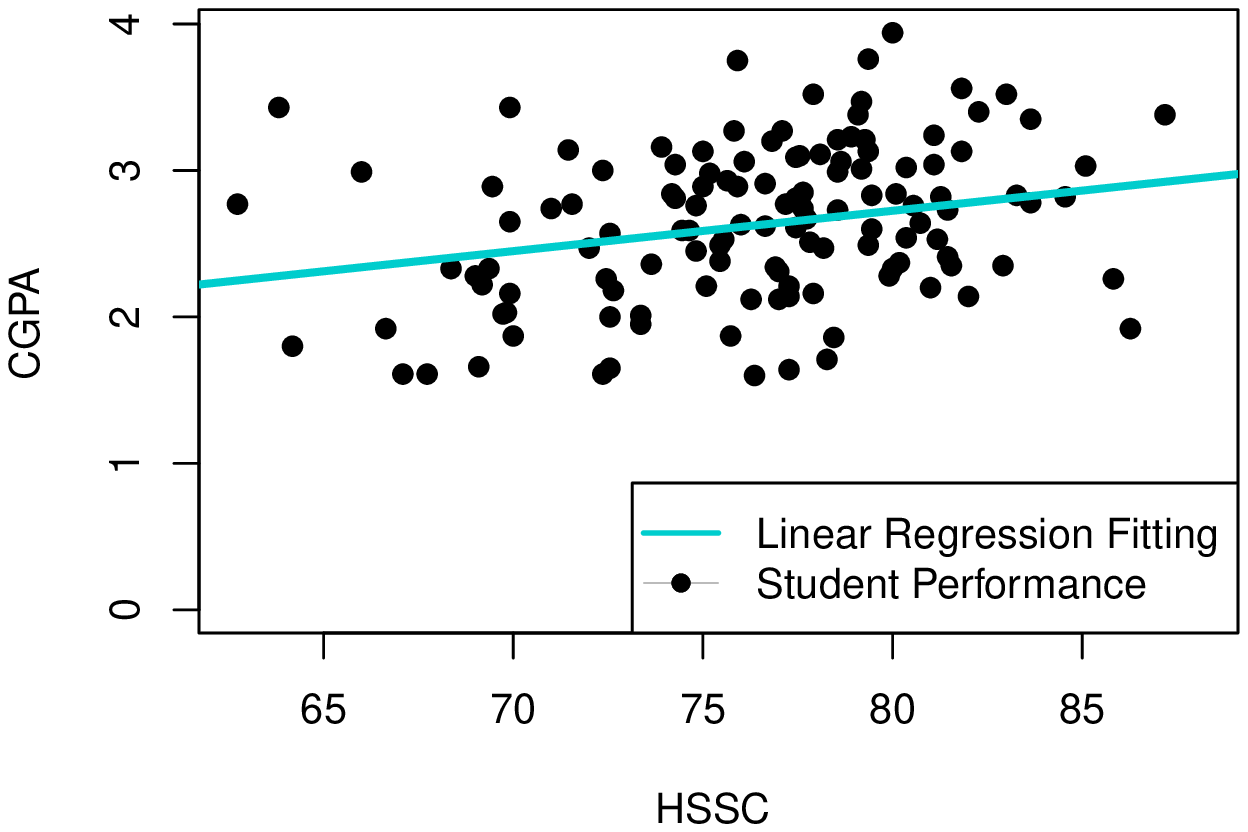}
\caption{Correlation between HSSC and CGPA}
\label{fig:linear-regression-fitting}
\end{minipage}
\end{figure*}

To find the pre-admission factors (SSC, HSSC, entry test and interview) that can predict student performance in the university Pearson Correlation has been applied. The result shows that there is a positive correlation between entry test and Cumulative Grade Point Average (CGPA) and also between HSSC and CGPA. The correlation coefficients ($r$) between the entry test and CGPA, and HSSC and CGPA are very close ($r$ = 0.29 and $r$ = 0.28 respectively), indicating that both entry test and HSSC are equally important in predicting the CGPA of a student. Figure \ref{fig:linear-regression-fitting-2} shows the correlation between the entry test of the students and their CGPA, and Figure \ref{fig:linear-regression-fitting} shows the correlation between the higher secondary school performance and the CGPA. These figures show that the students with a higher score in entry test and a higher percentage in HSSC performance obtain higher CGPA in the degree program.

\subsection{Grade Prediction}
For students, GPA prediction, students-courses matrix G is constructed. The data were transformed by centering the predictor variables by taking average GPA of a course and subtracted it from all GPAs of that course. 70\% of the dataset is used for training the CF MF and RBM models. Student GPAs for the courses has been predicted and displayed in Table \ref{table:gpa-predictions}.

\renewcommand{\arraystretch}{1.1}
\begin{table}[!h]
\caption{Student GPA prediction in courses based on CF, SVD, NMF and RBM technique}
\label{table:gpa-predictions}
\begin{tabularx}{\textwidth}{XXXXXXX}
\hline
Student Id. & Method & LCA & ECD & IS & SS & DL \\
\hline
SB145 & RBM & \textbf{2.67} & 3.67 & \textbf{2.33} & \textbf{3} & 4 \\
 & NMF & \textbf{1.86} & 3.67 & \textbf{1.99} & \textbf{3.61} & 4 \\    
 & SVD & \textbf{3.48} & 3.67 & \textbf{3.86} & \textbf{3.1} & 4 \\
 & UBCF & \textbf{2.99} & 3.67 & \textbf{2.99} & \textbf{2.91} & 4 \\
\hline
SB161 & RBM & \textbf{2.67} & 4 & \textbf{3} & \textbf{2.67} & 3.67 \\
 & NMF & \textbf{2.77} & 4 & \textbf{2.88} & \textbf{3.44} & 3.67 \\
 & SVD & \textbf{2.99} & 4 & \textbf{3.86} & \textbf{2.63} & 3.67 \\
 & UBCF & \textbf{2.41} & 4 & \textbf{2.81} & \textbf{2.39} & 3.67 \\
\hline
SB185 & RBM & \textbf{2.67} & 3.67 & \textbf{3} & \textbf{3.33} & \textbf{3} \\
 & NMF & \textbf{2.53} & 3.67 & \textbf{2.64} & \textbf{3.42} & \textbf{3.60} \\
 & SVD & \textbf{2.31} & 3.67 & \textbf{3.36} & \textbf{3.35} & \textbf{2.12} \\
 & UBCF & \textbf{1.84} & 3.67 & \textbf{2.51} & \textbf{2.93} & \textbf{2.12} \\
\hline
SB229 & RBM & \textbf{2.33} & \textbf{2} & \textbf{3.33} & \textbf{2} & \textbf{1.33} \\
 & NMF & \textbf{2.03} & \textbf{0.98} & \textbf{2.04} & \textbf{0.63} & \textbf{1.03} \\
 & SVD & \textbf{2.09} & \textbf{1.79} & \textbf{2.12} & \textbf{1.27} & \textbf{2.25} \\
 & UBCF & \textbf{2.77} & \textbf{2.19} & \textbf{3.14} & \textbf{1.42} & \textbf{2.43} \\
\hline
SB304 & RBM & 2 & \textbf{3} & 2.67 & \textbf{2} & \textbf{3} \\
 & NMF & 2 & \textbf{3.32} & 2.67 & \textbf{3.33} & \textbf{3.43} \\
 & SVD & 2 & \textbf{2.36} & 2.67 & \textbf{1.61} & \textbf{2.57} \\
 & UBCF & 2 & \textbf{2.19} & 2.67 & \textbf{1.42} & \textbf{2.43} \\
\hline
\end{tabularx}
\begin{tabular}{lll}
Predicted GPAs are in \textbf{bold} & Linear Circuit Analysis (LCA) &  Electronic Circuit \& Design (ECD) \\ Islamic Studies (IS) & Signals \& Systems (SS) & D-Lab (DL) \\
\end{tabular}
\end{table}

\subsection{Evaluation on model performance}
\begin{figure}[!h]
    \centering
    \includegraphics[width=0.73\textwidth]{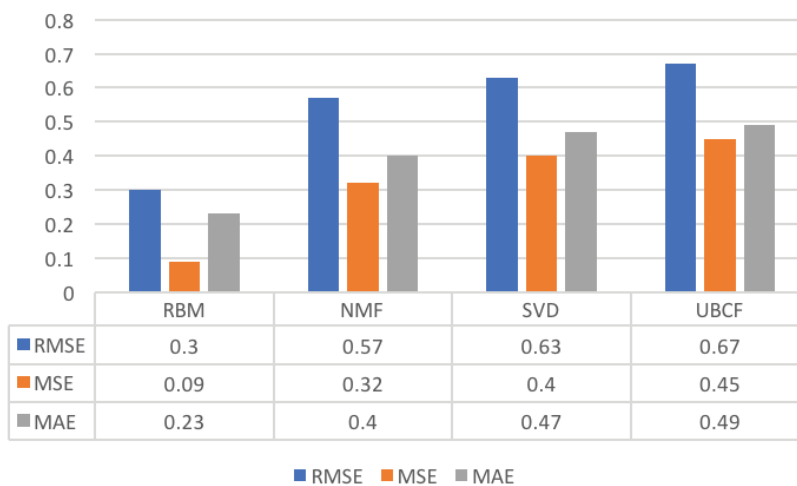}
    \caption{Evaluation of grade prediction models}
    \label{fig: error}
\end{figure}

There are several types of measures for evaluating the success of models. However, the evaluation of each model depends heavily on the domain and system's goals. For our system, our goal is to predict students' GPA and make decisions if a student needs to work hard to complete the course. These decisions work well when our predictions are accurate. To achieve it, we have to compare the prediction GPA against the actual GPA for the students-courses pair. Some of the most used metrics for evaluation of the models are the Root Mean Squared Error (RMSE), Mean Squared Error (MSE) and Mean Absolute Error (MAE). We evaluated model predictions by repeated random subsample cross-validation. We performed ten repetitions. In each run, we choose randomly 70\% of students data into the train set and 30\% of students data into the test set. We have computed RMSE, MSE, and MAE for each model. From Figure \ref{fig: error} the results show that the RBM model provides a clear improvement over the CF and MF models. Please note we are not performing student-level cross-validation of predicted results on newly registered students in this study but the currently enrolled students. 

\subsection{Feedback Model}

The results of feedback model that was discussed in detail in section 4 are shown in Table \ref{table:recommendation-results}. Here we put one of the students (SB185) to demonstrate the results of feedback model. We can see that the knowledge inference results of a student in Linear Circuit Analysis are less than 2.67, so the system gives a warning that the effort is needed in this course. These results are helpful for an instructor to identify weak students in a course by early predicting the grades and inferring student knowledge in the course domain.

\begin{table}[!h]
\centering
\caption{Feedback Model Result of Student (SB185)}
\label{table:recommendation-results}
\begin{tabular}{p{0.5in}p{0.6in}p{0.6in}p{1.5in}p{0.6in}p{0.7in}p{0.5in}}
  \hline
Course & Predicted Grade & Predicted GPA & Course Domain & Domain Average & Knowledge Inference & Effort Needed \\
  \hline
LCA & B & 3 & Electrical Engineering Foundation & 3.07 & 2.12 & YES \\
IS & B & 3 & Humanities & 3.19 & 3.83 & \\
SS & B- & 3.07 & Electrical Engineering Foundation & 3.4 & 2.91 & \\
DL & B+ & 3.33 & Management Sciences & 3.12 & 3.44 & \\
  \hline
\end{tabular}
\begin{tabular}{lll}
Linear Circuit Analysis (LCA) &  Electronic Circuit and Design (ECD) & Islamic Studies (IS) \\
Signals \& Systems (SS) & D-Lab (DL) & \\
\end{tabular}
\end{table}

\section{Insights} In this study, we have used CF (UBCF), MF (SVD and NMF) and RBM techniques to predict the students' performance in the courses. CF is a popular method to predict the students' performance due to its simplicity. In this technique, the students' performance is analyzed by using the previous data. It provides feedback to enhance the students' learning process based on the outcome of the analysis. However, this method has several disadvantages: since it depends upon the historical data of users or items for predicting the results. It shows poor performance when there is too much sparsity in the data, due to which we are not able to predict the students' performance accurately. Comparatively, in SVD technique, the data matrix $R$ is decomposed into users-features space and items-features space. When SVD technique is used with gradient descent algorithm to compute the best rank-k matrix approximation using only the known ratings of $R$, the accuracy of predicting the students' performance enhances but it may contain negative values which are hard to interpret. NMF technique enhances the meaningful interpretations of the possible hidden features that are obtained during matrix factorization. RBM is an unsupervised machine learning technique that is suitable for modeling tabular data. It provides efficient learning and inference better prediction accuracy than matrix factorization techniques. The use of RBM in recommender systems and e-commerce have also shown good results \cite{kanagal2012supercharging}. From the above discussion, it is clear that the RBM technique outperforms CF and MF techniques with lesser chances of error. The overall result obtained in this study also shows that RBM surpasses other techniques in predicting the student's performance.

\section{Limitations}
We note that the reported findings of this study have been based on the dataset of the performance of the undergraduate students from ITU. The dataset used in the study is limited with GPAs available for students in the particular courses. After using CF (UBCF), MF (SVD and NMF) and RBM techniques on the dataset, we can see that the RMSE for RBM technique is lower compared to the RMSE of other techniques. RMSE can be estimated with more clear results if more information of the students' GPAs is available. Student motivation during studies also plays a significant role in the prediction of student success which can be considered in future study related to the grade prediction. Moreover, there is a need to improve the prediction results by dealing with the cold-start problems. Also, models based on tensor factorization can be investigated to take the temporal effect into account in the student performance prediction. Despite these limitations, our research findings have important practical implications for the universities and institutes in enhancing their students' retention rate.

\section{Conclusion}
Early GPA predictions are a valuable source for determining student's performance in the university. In this study, we discussed CF (UBCF), MF and RBM techniques for predicting student's GPA. We use RBM machine learning technique for predicting student's performance in the courses. Empirical validation on real-world dataset shows the effectiveness of the used RBM technique. In a feedback model approach, we measure the students' knowledge in a particular course domain, which provides appropriate counseling to them about different courses in a particular domain by estimating the performance of other students in that course. This feedback model can be used as a component of an early warning system that will lead to students' motivation and provides them early warnings if they need to improve their knowledge in the courses. It also helps the course instructor to determine weak students in the class and to provide necessary interventions to improve their performance. In this way rate of the students' retention can be increased.

\bibliographystyle{acmtrans}
\bibliography{template}
\end{document}